\documentclass[12pt]{article}
\usepackage{amsfonts,amssymb,amsmath}
\usepackage[dvips]{epsfig}
\begin{document}
\title{\bf Signature Change in Noncommutative FRW Cosmology }
\author{T. Ghaneh$^1$\thanks{Email: ghaneh@tabrizu.ac.ir}
\hspace{2mm},  \hspace{2mm}
 F. Darabi$^2$\thanks{Email: f.darabi@azaruniv.edu (Corresponding author)   }
 \hspace{2mm}, and \hspace{2mm}
 H. Motavalli$^1$\thanks{Email: motavalli@tabrizu.ac.ir } \\
\centerline{$^1$\small {\em Department of Theoretical Physics
and Astrophysics, University of Tabriz, 51666-16471, Tabriz,
Iran.}}\\
{$^2$\small {\em Department of Physics, Azarbaijan Shahid Madani University, 53714-161, Tabriz, Iran. }}}

\maketitle
\begin{abstract}
\noindent
The conditions for which the no boundary proposal may have a classical realization
of a continuous change of signature, are investigated for a cosmological model described by FRW metric coupled with a self interacting scalar field, having a noncommutative phase space of dynamical variables. The model is then quantized and a good correspondence is shown between the classical and quantum cosmology indicating that the noncommutativity does not destruct the classical-quantum correspondence. It is also shown that the quantum cosmology supports a signature transition where the bare cosmological constant takes a vast continuous spectrum of negative values. The bounds of bare cosmological constant are limited by the values of noncommutative parameters. Moreover, it turns out that the physical parameters are constrained by the noncommutativity parametres.
\end{abstract}

\section{Introduction}

The application of Einstein's field equations to the system of universe always
faces with the problem of initial conditions. The big bang singularity is
such a well-known problem in the standard model of cosmology. However, one can remove this problem by presenting a physical realization for the philosophical concept of a universe with no beginning.
This presentation was firstly made by Hartle and Hawking in Ref.\cite{Hartle&Hawking}, where they showed that in the quantum interpretation of the very early
universe, it is not possible to express quantum amplitudes by 4-manifolds with globally Lorentzian geometries, instead they should be Euclidean compact manifolds with boundaries just located at a signature-changing hypersurface understood as the beginning of our Lorentzian universe.
This is well known as the {\it no boundary proposal}. In this direction of
thinking about quantum interpretation of the early universe, many works have
also been accomplished on different cosmological models to study whether it is possible to realize a classical signature change as an interpretation
of the no boundary proposal \cite{Dereli&Tucker,Dereli&Onder&Tucker,Darabi&Sepangi,Ghafoori&Gusheh&Sepangi,Vakili&Jalalzadeh&Sepangi,Ahmadi&Jalalzadeh&Sepangi}. Recently, the issue of signature change is again becoming of particular importance in the study of current acceleration of the universe \cite{Mars},
recovering the notion of time \cite{Jalal}, the challenge for quantum gravity \cite{Visser}, and the events in emergent spacetimes \cite{Visser1}.     However, no special attention has been paid for the cases where the recently
established notion of noncommutativity is applied to the phase space coordinates of the signature changing cosmological models. In this direction, an important question which may be considered as a problem is as follows: Does the quantum no boundary proposal have a classical realization representing a classical signature changing cosmology having variables in noncommutative phase space?

Abstract idea of noncommuting coordinates has been firstly proposed by Wigner
\cite{Wigner} for the thermodynamical phase space, and separately by Snyder \cite{Snyder} in offering an example of a Lorentz-invariant discrete space-time. The idea has been followed mathematically by Connes\cite{Connes} and Woronowicz \cite{Woronowicz} as noncommutative (NC) geometry, giving rise to a new formulation of quantum gravity through NC differential calculus \cite{Varilly,Madore}.
In another attempt, the link between NC geometry and string theory has become evident by the works of Seiberg and Witten \cite{Seiberg&Witten}, which resulted in NC field theories via the NC algebra based on the concept of Moyal product
\cite{Moyal,Szabo,Douglas}.

In this paper, we aim to study the effects of noncommutativity in the phase space of a cosmological model which exhibits the signature change in the commutative case. In section 2, we start with a FRW type metric and use a scalar field as the matter source of Einstein's field equations. Then, in
sections 3 and 4, we apply the noncommutativity to the phase space of the corresponding effective action by use of the Moyal product approach in deforming the Poisson bracket. The conditions for which the classical signature change is possible are then investigated. In section 5, we study the quantum cosmology of this noncommutative signature changing model and find the perturbative solutions for the corresponding Wheeler-DeWitt equation. Finally, in section
6, we pay attention to the interesting issue of classical limit. Since the idea of noncummutativity is relevant at early universe having quantum properties, it is important to investigate whether noncummutativity preserves the classical-quantum correspondence, namely the noncommutative quantum cosmology has a classical
limit.

\section{Classical Signature Dynamics in General Relativity}

We consider the same model which has already been extensively studied and
described by the metric \cite{Dereli&Tucker}
\begin{equation}\label{Eq1}
g=-\beta\, d\beta \otimes d\beta\ + \frac{\overline{R}^{2}(\beta)}{1+(k/4)r^{2}}\,(dx^{i} \otimes dx^{i}),
\end{equation}
in which $\overline{R}(\beta)$ is the scale factor of the universe, $k=-1,0,1$ determines the spatial curvature, and the hypersurface of signature change is identified by $\beta=0$.
Thus, the sign of $\beta$ is responsible for the geometry to be Lorentzian or Euclidian. The traditional cosmic time $t$ is related to $\beta$ via $t=\frac{2}{3}\beta^{3/2}$ when $\beta$ is definitely positive.
One way to treat the signature change problem is to find the exact solutions in Lorentzian region $(\beta>0)$ and then extrapolate them in Euclidian region. This kind of view assumes the Einstein's field equations to remain valid when passing through the $\beta=0$ junction. In Lorentzian domain, the line element (\ref{Eq1}) takes the form
\begin{equation}\label{Eq2}
ds^{2}=-dt^{2}+R^{2}(t)(dr^{2}+r^{2}d\Omega^{2}),
\end{equation}
where we have set $k=0$ in agreement with the current astronomical observations. We also assume the matter source to be an scalar field with interacting potential $U(\phi)$. The corresponding action
\begin{equation}
{\cal S}=\frac{1}{2\kappa^{2}}\int d^{4}x\sqrt{-g}{\cal R}+\int d^{4}x\sqrt{-g}\left[-\frac{1}{2}(\nabla\phi)^{2}-U(\phi)\right]+{\cal S}_{YGH},
\end{equation}
together with the metric (\ref{Eq2}) leads to the Lagrangian
\begin{equation}
{\cal L}=-3R\dot{R}^{2}+R^{3}\left[\frac{1}{2}\dot{\phi}^{2}-U(\phi)\right].
\end{equation}
Here, units are adopted so that $\kappa\equiv1$ and ${\cal S}_{YGH}$ is the York-Gibbons-Hawking boundary term. Note that a dot determines differentiation with respect to $t$. A change of dynamical variables $R,\phi\,(0\leq R<\infty,-\infty<\phi<+\infty)$ defined by
\begin{eqnarray}\label{6}
x_{1}=R^{3/2}cosh(\alpha\phi),
\end{eqnarray}
\begin{eqnarray}\label{7}
x_{2}=R^{3/2}sinh(\alpha\phi),
\end{eqnarray}
can express the Lagrangian in a more convenient form,
\begin{equation}
{\cal L}=\dot{x}_{1}^{2}-\dot{x}_{2}^{2}+ 2\alpha^{2}U(\phi)(x_{1}^{2}-x_{2}^{2}),
\end{equation}
where $\alpha^{2}=\frac{3}{8}$, and a coefficient ``$-2\alpha^{2}$" is neglected by using the zero energy condition\footnote{We know that general relativity is a {\it time reparametrization invariant}
theory. Every theory which is diffeomorphism invariant casts into the constraint
systems. Therefore, general relativity is a constraint system whose constraint
is the zero energy condition ${\cal H}=0$ \cite{Dirac,ADM}.}.
Following Ref.\cite{Dereli&Tucker}, we choose the potential $U(\phi)$ in a way that,
\begin{equation}\label{9}
2\alpha^{2}(x_{1}^{2}-x_{2}^{2})U(\phi)=a_{1}x_{1}^{2}+ a_{2}x_{2}^{2}+ 2b\,x_{1}x_{2},
\end{equation}
in which $a_{1},a_{2}$ and $b$ are constant parameters. Using (\ref{6}) and (\ref{7}), the equation (\ref{9}) implies
\begin{equation}\label{10}
U(\phi)=\lambda+\frac{1}{2\alpha^2}m^2 \sinh^2(\alpha \phi)+\frac{1}{2\alpha^2}b \sinh(2\alpha \phi),
\end{equation}
where the physical parameters
\begin{eqnarray}
&&\lambda=U\mid_{\phi=0}\,=a_{1}/2\alpha^{2},\\
&&m^{2}=\partial^{2}U/\partial\phi^{2}\mid_{\phi=0}\,=a_{1}+a_{2},
\end{eqnarray}
are defined as the cosmological constant and the mass of scalar field, respectively. The Hamiltonian of the system becomes
\begin{equation}
{\cal H}(x,p)=\frac{1}{4}(p_{1}^{2}-p_{2}^{2})-a_{1}x_{1}^{2}-a_{2}x_{2}^{2}-2b\,x_{1}x_{2},
\end{equation}
where $p_{1}, p_{2}$ are the momenta conjugate to $x_{1},x_{2}$, respectively.
The dynamical equations $\dot{x}_{i}=\{x_{i},{\cal H}\}, (i=1,2)$ then take the form \cite{Dereli&Tucker}
\begin{equation}
\ddot{\xi}={\sf M}\xi,
\end{equation}
where
\begin{equation}
{\sf M}=\left(
   \begin{array}{cc}
    a_{1} & b \\
    -b & -a_{2} \\
   \end{array}
  \right),\hspace{25mm}
\xi=\left(
        \begin{array}{c}
          x_{1} \\
          x_{2} \\
        \end{array}
      \right).
\end{equation}
Using the normal mode basis ${\sf V}={\sf S}^{-1}\xi=\left( \begin{array}{c} {\sf q}_{1}\\ {\sf q}_{2}\\ \end{array} \right)$ to diagonalize ${\sf M}$ as ${\sf S}^{-1}{\sf M}{\sf S}={\sf D}=diag({\sf m_{+}},{\sf m_{-}})$ we
find
\begin{equation}\label{Eq16}
{\sf m_{\pm}}=\frac{3\lambda}{4}-\frac{m^{2}}{2}\pm\frac{1}{2}\sqrt{m^{4}-4b^{2}},
\end{equation}
and the solutions obeying initial conditions $\dot{{\sf V}}(0)=0$ are obtained
as
\begin{equation*}
{\sf q}_{1}(t)=2{\sf A}_{1}\cosh(\sqrt{{\sf m_{+}}}\,t),
\end{equation*}
\begin{equation}
{\sf q}_{2}(t)=2{\sf A}_{2}\cosh(\sqrt{{\sf m_{-}}}\,t),
\end{equation}
where ${\sf A}_{1},{\sf A}_{2}\in\mathbb{R}$. These solutions remain real when the phase of $(\sqrt{{\sf m_{+}}}\,t)$ changes by $\pi/2$, so they
are candidates for determining real signature changing geometries. However, the constants ${\sf A}_{1}$ and ${\sf A}_{2}$ are correlated by the zero energy condition
\cite{Dereli&Tucker}
\begin{equation}\label{V}
V^T(0)\mathcal I V(0)=0,
\end{equation}
where $\mathcal I={\sf S}^T {\sf J}{\sf M}{\sf S}$ and
$$ {{\sf J}=\left(\begin{array}{cc}
    1 & 0 \\
    0 & -1 \\
   \end{array}
  \right).}
$$
Eq.(\ref{V}) is a quadratic equation for the ratio $\chi={\sf A}_{1}/{\sf A}_{2}$ and has has roots $\chi_{\pm}$ determined in terms of the parameters of $\lambda, m^2,b$. By choosing an overall scale through ${\sf A}_{2} =1$, the solutions fall into two classes
\begin{equation}\label{}
\xi^{\pm}(t)={\sf S}{\sf V}^{\pm}(t),
\end{equation}
where
\begin{equation}\label{}
{\sf q}^{\pm}_1(t)=2{\sf A}_{1}^{\pm}\cosh(\sqrt{{\sf m_{+}}}\,t),
\end{equation}
and
\begin{equation}\label{}
{\sf q}^{\pm}_2(t)=2\cosh(\sqrt{{\sf m_{-}}}\,t).
\end{equation}
Finally, $R$ and $\phi$ are recovered from $x_{1}$ and $x_{2}$ via (\ref{6}) and (\ref{7}) as
\begin{equation}\label{}
R(t)=(x_{1}^2-x_{2}^2)^{1/3},
\end{equation}
\begin{equation}\label{}
\phi(t)=\frac{1}{\alpha}\tanh^{-1}\left(\frac{x_{2}}{x_{1}}\right).
\end{equation}

It turns out that: i) if both eigenvalues of {\sf M} are positive, then no signature transition occurs, ii) with the product of the eigenvalues less than zero, the constraint (\ref{V}) cannot be satisfied with a real solution for the amplitude $\chi$, and iii) with both eigenvalues negative, then typically $x_{1}(\beta), x_{2}(\beta)$ exhibit bounded oscillations in the region $\beta> 0$ and are unbounded for $\beta< 0$ (see figure 1 in \cite{Dereli&Tucker}). Such behaviour translates into the solutions for $R$ and $\phi$ (see figure 2 in \cite{Dereli&Tucker}). Thus it is possible to choose parameters so that the metric has Euclidean signature for a finite range of $\beta< 0$. It undergoes a transition at $\beta= 0$ to Lorentzian signature, afterwards it persists for a further finite range of $\beta> 0$ \cite{Dereli&Tucker}.

\section{Classical Noncommutativity}

A common approach to study the noncommutativity between phase space variables is based on replacing the usual product between the variables with the so called star-product.
For flat Euclidian spaces, all the star-products are \emph{c-equivalent} to the so called Moyal product \cite{Hirshfeld}.\\
Suppose $f(x_{1},..,x_{n};p_{1},..,p_{n})\,,g(x_{1},..,x_{n};p_{1},..,p_{n})$ are two arbitrary functions. The Moyal product is defined as
\begin{equation}\label{Eq17}
f\star_{\propto}g=f\,e^{\frac{1}{2}\overleftarrow{\partial}_{a}\propto_{ab}\overrightarrow{\partial}_{b}}g,
\end{equation}
such that
\begin{equation}
\propto_{ab}=\left(
               \begin{array}{cc}
                 \theta_{ij} & \delta_{ij}+\sigma_{ij} \\
                 -\delta_{ij}-\sigma_{ij} &\bar{\theta}_{ij} \\
               \end{array}
             \right),
\end{equation}
and $\theta_{ij},\bar{\theta}_{ij}$ are antisymmetric $N\times N$ matrices.
The deformed Poisson brackets then reads
\begin{equation}
\{f,g\}_{\propto}=f\star_{\propto}g-g\star_{\propto}f.
\end{equation}
Hence, coordinates of a phase space equipped with Moyal product satisfy
\begin{equation}\label{Eq20}
\{x_{i},x_{j}\}_{\propto}=\theta_{ij},\hspace{20mm} \{x_{i},p_{j}\}_{\propto}=\delta_{ij}+\sigma_{ij},\hspace{20mm} \{p_{i},p_{j}\}_{\propto}=\bar{\theta}_{ij}.
\end{equation}
On the other hand, considering the following transformations \cite{Masud}-\cite{Masud2}
\begin{equation}\label{Eq21}
x^{\prime}_{i}=x_{i}-\frac{1}{2}\theta_{ij}p^{j},\hspace{20mm} p^{\prime}_{i}=p_{i}+\frac{1}{2}\bar{\theta}_{ij}x^{j},
\end{equation}
$(x^{\prime}_{i},p^{\prime}_{j})$ fulfill the same commutation relations as (\ref{Eq20}), but with respect to the usual Poisson brackets
\begin{equation}
\{x^{\prime}_{i},x^{\prime}_{j}\}=\theta_{ij},\hspace{20mm} \{x^{\prime}_{i},p^{\prime}_{j}\}=\delta_{ij}+\sigma_{ij},\hspace{20mm} \{p^{\prime}_{i},p^{\prime}_{j}\}=\bar{\theta}_{ij},
\end{equation}
provided that $(x_{i},p_{j})$ follows the common commutation relations
\begin{equation}
\{x_{i},x_{j}\}=0,\hspace{20mm} \{p_{i},p_{j}\}=0,\hspace{20mm} \{x_{i},p_{j}\}=\delta_{ij}.
\end{equation}
The latter approach is called \emph{noncommutativity via deformation}.

\section{Signature Change in Noncommutative Phase Space}

 In the previous section we mentioned that equipping the phase space with the noncommutative Moyal product (\ref{Eq17}) is equal to applying the transformations (\ref{Eq21}) to the phase space coordinates. This technique makes the definition of noncommutative Hamiltonian easy as follows
\begin{equation}\label{Eq23}
{\cal H}^{\prime}(x^{\prime},p^{\prime}) =\frac{1}{4}(p^{\prime\,2}_{1}-p^{\prime\,2}_{2})-a_{1}x^{\prime\,2}_{1}-a_{2}x^{\prime\,2}_{2}-2bx^{\prime\,1}x^{\prime\,2}.
\end{equation}
In this two-dimensional case, $\theta,\bar{\theta}$ and $\sigma$ have simple forms \cite{Djemai&Smail}: \begin{equation}\label{Eq24}
\theta_{ij}=\theta\epsilon_{ij},\hspace{20mm}\bar{\theta}_{ij}=\bar{\theta}\epsilon_{ij}, \hspace{20mm}\sigma_{ij}=\sigma\epsilon_{ij},\hspace{20mm} \sigma=\frac{1}{4}\theta\bar{\theta},
\end{equation}
with $\epsilon_{ij}$ being the totally anti-symmetric tensor. The map (\ref{Eq21}),
regarding (\ref{Eq24}), converts the noncommutative Hamiltonian (\ref{Eq23}) into a new commutative one
\begin{equation}\label{Eq25}
{\cal H}^{\prime}(x,p)=\frac{1}{4}(b_{2}p_{1}^{2}-b_{1}p_{2}^{2})-c_{1}x_{1}^{2}-c_{2}x_{2}^{2} +d_{1}x_{1}p_{2}+d_{2}x_{2}p_{1}-2bx_{1}x_{2}+\frac{1}{2}b\theta^{2}p_{1}p_{2}-b\theta(x_{1}p_{1}-x_{2}p_{2}),
\end{equation}
where $x_{i},p_{j}$ reads the common Poisson algebra, and
\begin{eqnarray}
\nonumber && b_{1}=1+\theta^{2}a_{1},\hspace{26mm} b_{2}=1-\theta^{2}a_{2}, \\
\nonumber && c_{1}=a_{1}+(\bar{\theta}/4)^{2},\hspace{21mm} c_{2}=a_{2}-(\bar{\theta}/4)^{2}, \\
&& d_{1}=(\bar{\theta}/4)+\theta a_{1},\hspace{20mm} d_{2}=(\bar{\theta}/4)-\theta a_{2}.
\end{eqnarray}
One can write the classical equations of motion $\dot{x}_{i}=\{x_{i},{\cal H}^{\prime}\}$ in the matrix form
\begin{equation}\label{Eq27}
c\, \ddot{\xi}(t)+N\,\dot{\xi}(t)+M\,\xi(t)=0,
\end{equation}
with
\begin{equation}
\xi=\left(
  \begin{array}{c}
    x_{1} \\
    x_{2} \\
  \end{array}
\right),\hspace{7mm}
N=\left(
  \begin{array}{cc}
    n_{11} & n_{12} \\
    n_{21} & n_{22} \\
  \end{array}
\right),\hspace{5mm}
M=\left(
  \begin{array}{cc}
    m_{11} & m_{12} \\
    m_{21} & m_{22}, \\
  \end{array}
\right),
\end{equation}
and
\begin{eqnarray}
\nonumber && n_{11}=-n_{22}=-b \theta l , \\
\nonumber && n_{12}=b_{2}l/\theta , \hspace{5mm} n_{21}=b_{1}l/\theta , \\
\nonumber && m_{11}=-(\sigma-1)^{2}\left[e(b_{2}+1)\theta^{2}-a_{1}\right] , \\
\nonumber && m_{22}=+(\sigma-1)^{2}\left[e(b_{1}+1)\theta^{2}+a_{2}\right] , \\
          && m_{12}=-m_{21}=b(\sigma-1)^{2}(e\theta^{4}+1) ,
\end{eqnarray}
where
\begin{eqnarray}
\nonumber && c=(a_{2}-a_{1})\theta^{2}+e\theta^{4}-1 , \\
\nonumber && l=2(\sigma-e\theta^{4})-(a_{2}-a_{1})(\sigma+1)\theta^{2} , \\
          && e=a_{1}a_{2}-b^{2}.
\end{eqnarray}

Now, we solve Eq.(\ref{Eq27}) by picking a normal mode basis to diagonalize $M$ and $N$. These are simultaneously diagonalizable by the matrix $S$ if one of the following three conditions is satisfied \footnote{ Actually these conditions are somehow the necessary conditions for $R$ and $\phi$ to be real: If $x_{1},x_{2}$ can not be decoupled, then $x_{1}$ will remain related to $\dot{x}_{2}$ (and also $x_{2}$ related to $\dot{x}_{1}$), which means both $x_{1}, x_{2}$ can not satisfy $\dot{x}_{i}=0$, hence
can not be as $cosh$ functions, simultaneously. This results in non-real valued $R$ or $\phi$ in $\beta<0$ region. }

\begin{equation}\label{Eq29}
 l=0 , \hspace{10mm} e\theta^{4}+1=0, \hspace{10mm} m^2=b=0.
\end{equation}

Each of the last two choices in (\ref{Eq29}) leads to an infinite scale factor or scalar field. This leaves us only with the first case, namely $l=0$, to
proceed. The diagonalization process then gives

\begin{equation}\label{Eq30}
S^{-1}MS=D=\left(
  \begin{array}{cc}
    m_{+} & 0 \\
    0 & m_{-} \\
  \end{array}
\right), \hspace{15mm} S^{-1}NS=0 .
\end{equation}
By defining
\begin{equation}\label{Eq31a}
\xi(t)=S\,V(t),\hspace{25mm}
V(t)=\left(
    \begin{array}{c}
      q_{1}(t) \\
      q_{2}(t) \\
    \end{array}
   \right),
\end{equation}
where
\begin{equation}\label{Eq32}
S=\left(
   \begin{array}{cc}
    m_{12}/s_{+} & m_{12}/s_{-} \\
         1       &      1       \\
   \end{array}
  \right),
\end{equation}
and
\begin{eqnarray}
\nonumber && m_{\pm}=\frac{1}{2}(m_{22}+m_{11}\pm\sqrt{\Delta}),\\
&& s_{\pm}=\frac{1}{2}(m_{22}-m_{11}\pm\sqrt{\Delta}),\\
&& \Delta=(m_{22}-m_{11})^{2}+4m_{12}m_{21},
\end{eqnarray}
the coupled equations (\ref{Eq27}) are converted into the following decoupled
equations
\begin{equation}
\ddot{V}(t)=DV(t),
\end{equation}
having the general solution
\begin{equation}
V(t)=\Lambda_{+}(t)\textbf{A}_{+}+\Lambda_{-}(t)\textbf{A}_{-},
\end{equation}
with $\textbf{A}_{+},\textbf{A}_{-}$ being constant vectors, and
\begin{equation}
\Lambda_{\pm}=
\left(
   \begin{array}{cc}
    e^{\pm i\sqrt{\frac{m_{+}}{c}}\,t} & 0 \\
         0       &      e^{\pm i\sqrt{\frac{m_{-}}{c}}\,t}  \\
   \end{array}
\right).
\end{equation}
Requesting for the initial conditions $\dot{V}(0)=0$, implies that \footnote{Demanding real-valued solutions in Lorentzian region, we find that the initial conditions
$\dot{V}(0)=0$ guarantee the solutions to remain real when passing through the hypersurface of signature change toward the Euclidean area.}
\begin{equation}\label{Eq37}
q_{1}(t)=A_{1}\,\cosh\left(i\sqrt{\frac{m_{+}}{c}}\,t\right),\hspace{20mm} q_{2}(t)=A_{2}\,\cosh\left(i\sqrt{\frac{m_{-}}{c}}\,t\right),
\end{equation}
where $A_{1},A_{2}, \in {\mathbb R}$.
Then, $x_{1},x_{2}$ are immediately obtained from (\ref{Eq31a}) as

\begin{equation}\label{Eq38}
x_{1}(t)=m_{12}\left(\frac{q_{1}(t)}{s_{+}}+\frac{q_{2}(t)}{s_{-}}\right) ,\hspace{20mm}
x_{2}(t)=q_{1}(t)+q_{2}(t),
\end{equation}
and one can also find $p_{1},p_{2}$ by using the dynamical equations as
follows
\begin{eqnarray}
p_{1}(t)=2\left[b\theta^{2}\dot{x}_{2}(t)+b_{1}\dot{x}_{1}(t)+b\theta(b_{1}-\theta d_{1})x_{1}(t)-(d_{2}b_{1}+b^{2}\theta^{3})x_{2}(t)\right]/(b^{2}\theta^{4}+b_{1}b_{2}) ,
\end{eqnarray}
\begin{eqnarray}
 p_{2}(t)=2\left[b\theta^{2}\dot{x}_{1}(t)-b_{2}\dot{x}_{2}(t)+b\theta(b_{2}-\theta d_{2})x_{2}(t)+(d_{1}b_{2}+b^{2}\theta^{3})x_{1}(t)\right]/(b^{2}\theta^{4}+b_{1}b_{2}).
\end{eqnarray}

Regarding (\ref{Eq37}), (\ref{Eq38}), the above results show that the momentum
fields contain $sinh$ functions and therefore are imaginary in Euclidean area. This asserts the junction condition $\dot{x_i}(0)=0$. In general, the junction condition, apart from continuity of the fields, is that the momenta conjugate to the fields must vanish on the hypersurface of signature change. This junction condition on signature-changing solutions, is also referred to as real tunnelling solutions in the context of quantum cosmology, and the familiar argument is that the momentum fields are real in the Lorentzian region and imaginary in the Riemannian region, hence must vanish at the junction \cite{Hayward}.

The results (\ref{Eq37}) are the solutions of dynamical equations in Lorentzian area. These solutions must be bounded as ``$\cos$" functions in $\beta>0$ region which requires $m_{\pm}/(-c)$ to be negative. Calculations indicate that this requirement can not be reached for a zero $b$ which shows the crucial role of the cross-term $bx_{1}x_{2}$ in the Hamiltonian for the signature change. This is in agreement with the claim in Ref.\cite{Dereli&Tucker} in
that the presence of the cross term breaks the symmetry of $U(\phi)$ under $\phi \rightarrow -\phi$ and is directly responsible for the signature changing properties of the solutions. Choosing a bigger allowed value of $b$ leads $m_{+}$ , $m_{-}$ , $s_{+}$ , $s_{-}$ and also $A_{1}, A_{2}$ to be more close,
in order of magnitude, to each other. Another result coming from $m_{\pm}/c>0$ is that the allowed values of $\theta, \bar{\theta}$ are of the same sign, i.e. $\sigma>0$. Trivial solutions for $R$ and $\phi$ are obtained when $a_{1}=a_{2}=\pm b$ or when $b_{1}+b_{2}=0$, $a_{2}\pm b=1/\theta^{2}$.

Figures 1 to 3 show the signature change from Euclidean to Lorentzian in the sense of continuous transition of physical variables from $\beta<0$ to $\beta>0$ regions, according to \cite{Dereli&Tucker} \footnote{In these figures, the values of $\theta, \bar{\theta}, \lambda$ and $b$ are finely selected in order to satisfy the equation(\ref{Eq51}), $m_{\pm}/c>0$ and the conditions ${\cal H}=0$ and ${\cal R}\mid_{_{\beta=0}}=0$. It is also worth to note that changing the order of magnitude of these parameters does not affect drastically the shape and physical behavior of these plots.}. As is evident, the roots of $\bar{R}(\beta)$ admits the singularities of both $\bar{\phi}(\beta)$ and $\bar{\cal{R}}(\beta)$.

\begin{figure}
  \includegraphics[width=8.5cm]{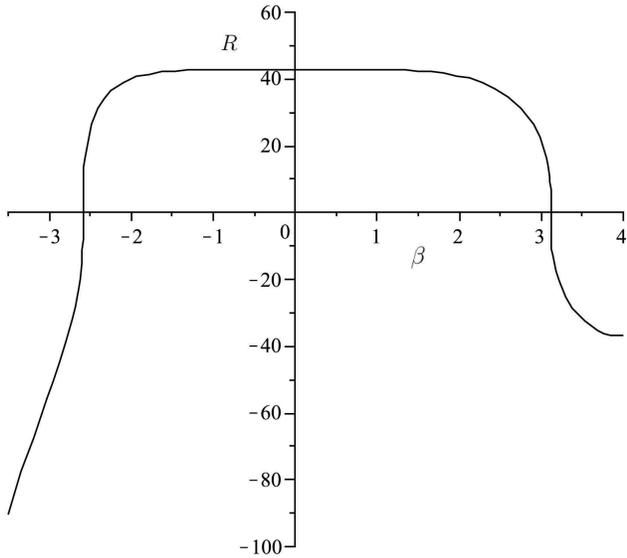}\\
  \caption{ {\small The behavior of scale factor with respect to $\beta$ for $\theta=2.6\times10^{-5},~ \bar{\theta}=2.0\times10^{-5},~\lambda=0.193$ and $b=5\times10^{-4}$.
The upper half plane $R\geqslant0$ is physically viable.  } } \label{Fig1}
\end{figure}

\begin{figure}
  \includegraphics[width=8.5cm]{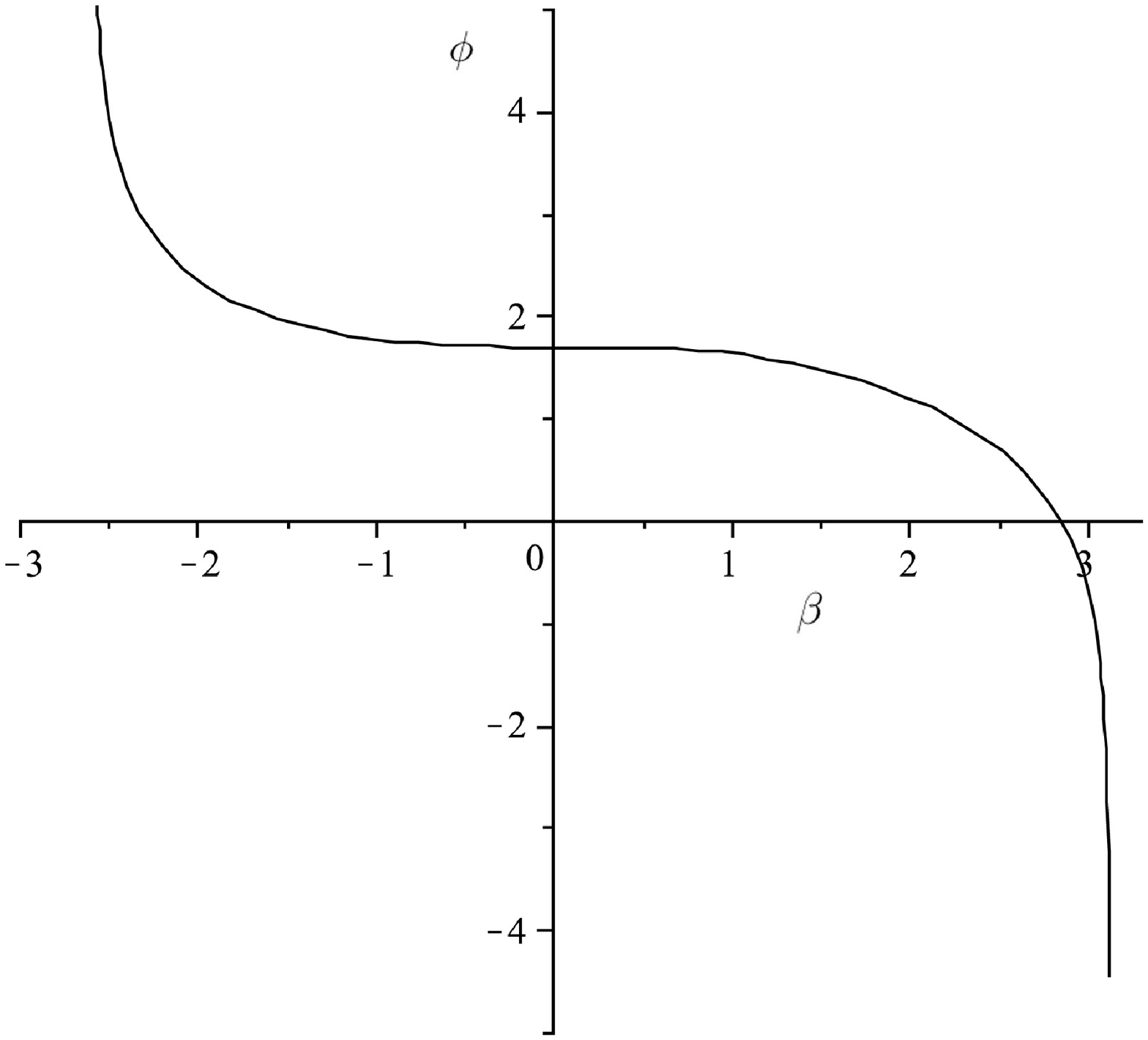}\\
  \caption{ {\small The behavior of scalar field with respect to $\beta$ for $\theta=2.6\times10^{-5},~ \bar{\theta}=2.0\times10^{-5},~\lambda=0.193$ and $b=5\times10^{-4}$.} } \label{Fig2}
\end{figure}

\begin{figure}
  \includegraphics[width=8.5cm]{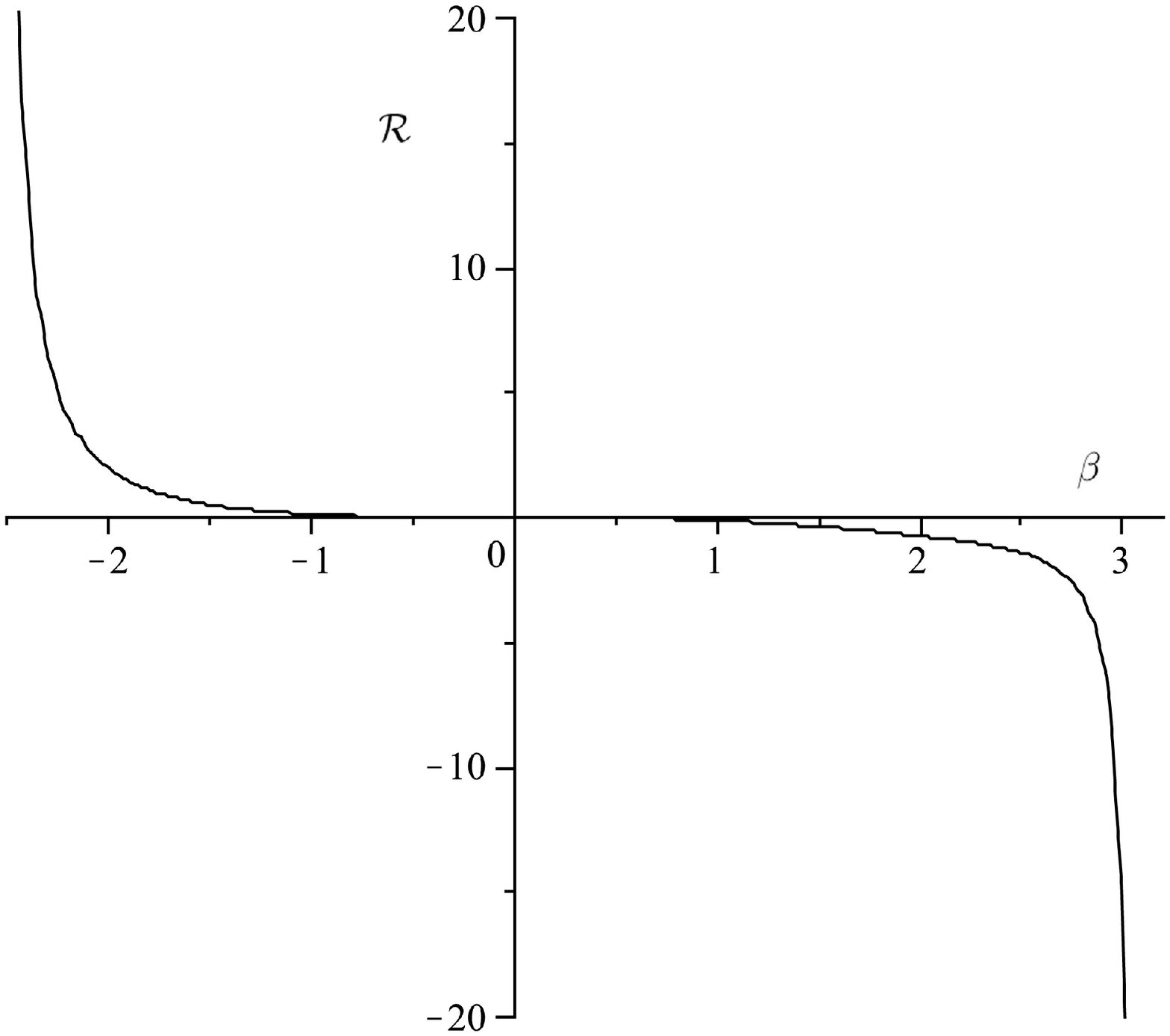}\\
  \caption{ {\small The behavior of Ricci scalar with respect to $\beta$ for $\theta=2.6\times10^{-5},~ \bar{\theta}=2.0\times10^{-5},~\lambda=0.193$ and $b=5\times10^{-4}$.} } \label{Fig3}
\end{figure}

\section{Quantum Cosmology}

Clearly, in high energy physics, where energies are of the order of Planck mass, we need an essential revision in the concept of space-time. One such revision is to consider noncommutativity in the phase space. The early universe with ultra high energy and a very small size of Planck length is the most appropriate {\it quantum laboratory} which is hoped to give useful information in this concern. So, it is interesting to study the quantized version of the present classical noncommutative signature changing model and also check for the important issue of classical-quantum correspondence.

Applying noncommutativity to a quantized model is usually studied perturbatively.
This urges $\sigma$ as a perturbative parameter to satisfy\footnote{Note that $\sigma=1$ is a singularity of the scalar field, hence is forbidden.}  $\sigma^{2}<1$.
Meanwhile, we consider $b$ and $m^{2}$ as infinitesimal parameters. The quantized hamiltonian then becomes
\begin{equation}\label{WD}
\hat{{\cal H}}^{\prime}=\hat{{\cal H}}^{\prime}_{0} + \hat{{\cal V}}^{\prime},
\end{equation}
where a coefficient of $b_{1}/4$ is omitted according to the zero energy condition and
\begin{eqnarray}
\nonumber&& \hat{{\cal H}}^{\prime}_{0}=\hat{H}_{1}-\hat{H}_{2}+\frac{4}{b_{1}}d_{1}\hat{L}_{12} ,\\
&& \hat{{\cal V}}^{\prime} = \frac{4}{b_{1}}\left(b\,\hat{\textsf{V}}_{1}-m^{2}\,\hat{\textsf{V}}_{2}\right),
\end{eqnarray}
where
\begin{eqnarray}
 &&\nonumber \hat{H}_{i}=\hat{p}_{i}^{2}+\Omega_{i}^{2}\hat{x}_{i}^{2},\hspace{20mm} i=1,2 , \\
 &&\nonumber \hat{L}_{12}=\hat{x}_{1}\hat{p}_{2}+\hat{x}_{2}\hat{p}_{1}, \\
 && \Omega^{2}=-4c_{1}/b_{1}, \\
 &&\nonumber \hat{\textsf{V}}_{1}=-2\hat{x}_{1}\hat{x}_{2}+\frac{\theta^{2}}{2}\hat{p}_{1}\hat{p}_{2}-\theta(\hat{x}_{1}\hat{p}_{1}-\hat{x}_{2}\hat{p}_{2}), \\
 && \hat{\textsf{V}}_{2}=(\frac{\theta}{2}\hat{p}_{1}+\hat{x}_{2})^{2}.
\end{eqnarray}
The assumption $\sigma^{2}<1$ and solvability condition $\Omega^{2}\geq0$ are just fulfilled provided that

\begin{equation}\label{Eq51}
-4/3\theta^{2}\leq\lambda\leq-\bar{\theta}^{2}/12.
\end{equation}
At first, we try to find the eigenfunctions of the non-perturbed Hamiltonian $\hat{{\cal H}}^{\prime}_{0}$. Let us define a new set of variables $(\rho,\varphi)$ by \cite{Sepangi}
\begin{equation}
 x_{1}=\rho\cosh\varphi,\hspace{25mm}x_{2}=\rho\sinh\varphi.
\end{equation}
Then, we obtain the quantum operators by using the common rule
$p_q \rightarrow-i\partial/\partial q$ as
\begin{eqnarray}
&& L_{12}=-i(x_{1}\frac{\partial}{\partial x_{2}}+x_{2}\frac{\partial}{\partial x_{1}})=-i\frac{\partial}{\partial\varphi}, \\
&& \hat{H_{1}}-\hat{H_{2}}=\frac{\partial^{2}}{\partial x_{2}^{2}}-\frac{\partial^{2}}{\partial x_{1}^{2}}+\Omega^{2}(x_{1}^{2}-x_{2}^{2})=-(\frac{\partial^{2}}{\partial\rho^{2}}+ \frac{1}{\rho}\frac{\partial}{\partial\rho})+\Omega^{2}\rho^{2}.
\end{eqnarray}

A function of the form $\psi(\rho,\varphi)={\cal U}(\rho)e^{-in\varphi}$ is an eigenfunction of $L_{12}$ with eigenvalue $n$. It is easy to show that $L_{12}$ commutes with $\hat{H_{1}}-\hat{H_{2}}$ which means $\psi(\rho,\varphi)$ can be an eigenfunction of the whole $\hat{{\cal H}}^{\prime}_{0}$. This choice of solution separates the non-perturbed Wheeler-DeWitt equation (\ref{WD})
and gives the following differential equation for ${\cal U}(\rho)$
\begin{equation}
\frac{d^{2}\,{\cal U}}{d\rho^{2}}+\frac{1}{\rho}\frac{d\,{\cal U}}{d\rho}+\left(\frac{\nu^{2}}{\rho^{2}}-\Omega^{2}\rho^{2}-4\nu\frac{d_{1}}{b_{1}}\right)\,{\cal U}=0.
\end{equation}
Applying a change of variable $r=2\Omega\rho^{2}$ and a transformation ${\cal U}=\frac{1}{\rho}{\cal W}$ yields the Whittaker differential equation
\begin{equation}
\frac{d^{2}{\cal W}}{dr^{2}}+\left(\frac{-1}{4}+\frac{\kappa}{r}+\frac{\frac{1}{4}-\nu^{2}}{r^{2}}\right){\cal W}=0,
\end{equation}
where $\nu=in/2$ and $\kappa=\mp n d_{1}/\Omega b_{1}$. This equation has a solution which can be expressed in terms of confluent hypergeometric functions $M(a,b;x)$ and $U(a,b;x)$ as
\begin{equation}
{\cal W}(r)=e^{-r/2}r^{\nu+\frac{1}{2}} \left[c\,U(\nu-\kappa+\frac{1}{2},2\nu+1;r)+c^{\prime}\,M(\nu-\kappa+\frac{1}{2},2\nu+1;r)\right].
\end{equation}
Following \cite{Sepangi}, we set $c^{\prime}=0$ due to asymptotic behavior of $M(a,b;x)\sim e^{x}/x^{b-a}$\cite{Abramowitz}. Hence, the eigenfunctions of $\hat{{\cal H}}^{\prime}_{0}$ becomes

\begin{equation}
\psi_{n}^{\pm}(\rho,\varphi)=\rho^{in}e^{\mp\Omega\rho^{2}/2}\,U\left(\frac{in+1}{2}\pm n d_{1}/\Omega b_{1},in+1;\pm\Omega\rho^{2}\right)e^{-in\varphi},
\end{equation}
among which only $\psi_{n}^{+}$ are normalizable. Note that $n$ should be an integer in order to $\psi_{n}^{+}$ be single-valued functions of $\varphi$,
so one can at last write the general non-perturbed wave function as
\begin{equation}
\psi(\rho,\varphi)=\sum\limits_{n=-\infty}^{+\infty}c_{n}\rho^{in}e^{-\Omega\rho^{2}/2}\, U\left(\frac{in+1}{2}+n d_{1}/\Omega b_{1},in+1;\Omega\rho^{2}\right)e^{-in\varphi}.
\end{equation}

Now, let us consider the system perturbed by $\hat{{\cal V}}^{\prime}$. According to the time-independent perturbation theory to the first order, the perturbed part of the solution, which is named here by $\chi$, is the eigenfunction of $\hat{{\cal V}}^{\prime}$, namely
\begin{equation}\label{Eq57}
\hat{{\cal V}}^{\prime} \chi = 0.
\end{equation}
A general solution of $\textsf{V}_{1}\chi_{1}=0$ which has equal real and imaginary parts is

\begin{equation}\label{Eq58}
\chi_{1}(x_{1},x_{2})=(1+i) \left[c^{\prime}_{1}\,J(0,\frac{2}{\theta}x_{1}x_{2})+c^{\prime}_{2}\,Y(0,\frac{2}{\theta}x_{1}x_{2})\right],
\end{equation}
where $J$ and $Y$ are ordinary \textit{Bessel} functions of the first and second kind, respectively.
Also the general solution of $\textsf{V}_{2}\chi_{2}=0$ is
\begin{equation}\label{Eq59}
\chi_{2}(x_{1},x{2})=F(x_{2})e^{\frac{2}{\theta}x_{1}x_{2}},
\end{equation}
$F(x_{2})$ being an arbitrary complex function.
Now, one can determine $F(x_{2})$ in a way that $\chi(x_{2})$ also satisfies $\textsf{V}_{1}\chi_{2}=0$, thus $\chi_{2}$ can be a solution of (\ref{Eq57}).
So, regarding $\chi=b\chi_{1}+m^{2}\chi_{2}$, the wave function of the perturbed universe in the first order is finally obtained as
\begin{equation}\label{WF}
\Psi(\rho,\varphi)=\psi(\rho,\varphi)+b\chi_{1}(\rho,\varphi)+m^{2}\chi_{2}(\rho,\varphi).
\end{equation}

\section{Classical limit}

One of the interesting topics in the context of quantum cosmology is the
classical limit, namely finding the mechanisms by which classical cosmology may emerge from quantum cosmology. In other words, how the wavefunction of universe predicts a classical spacetime? Most authors consider semiclassical approximations to the Wheeler DeWitt equation and refer to regions in configuration space where the solutions of Wheeler DeWitt equation are oscillatory or exponentially decaying. The
former represents classically allowed region while the latter represents the forbidden region. These regions are determined by the initial conditions
imposed on the wave function. Two popular proposals for the initial conditions are the {\it no boundary} \cite{Haw-Haw1} and {\it tunneling} \cite{Vil-Vil1} proposals. The idea of classical signature change has its origin in the no boundary
proposal and that is why we are interested in the classical-quantum correspondence
to characterize the classical signature change as the classical limit of
no boundary proposal.

In general, the quantum states do not offer semiclassical description of some spacetime domain unless one introduces a decoherence mechanism widely regarded as necessary to assign a probability for the occurrence of a classical metric. However, in order for a simple and satisfactory classical-quantum correspondence is achieved in the lack of decoherence mechanism, we may investigate if the absolute values of the solutions of Wheeler-DeWitt equation have maxima in the vicinity of the classical loci. In fact, a viable quantization should be one that has good chance of yielding a classical limit not too far from the classical predictions. This line of thought has already been extensively
pursued by many authors and good classical-quantum correspondences are obtained \cite{Kiefer}-cite{Kiefer2}. Following this point of view for studying the classical limit
we investigate the classical-quantum correspondences in the present noncommutative
model. In figure 4, the classical loci (\ref{Eq37}) and (\ref{Eq38}), and the density plot of the wavefunctions (\ref{WF}) are superimposed for small values of $b$ and $m^{2}$. A good correspondence is seen between the classical and quantum cosmology. The point is that, in general, it seems the presence of noncommutative parameters may allow us to achieve better correspondence than commutative case, between the classical and quantum cosmology. This
is because, in the commutative case we have just three parameters ($\lambda, b, m^2$) as degrees of freedom while in the noncommutative case we have a
set of enlarged parameters ($\theta, \bar{\theta}, \lambda, b, m^2$) having
one more degree of freedom (we have a constraint (\ref{m})), so one may achieve
a better classical-quantum  correspondence by adjusting four rather than three degrees of freedom.

\begin{figure}
  \includegraphics[width=8.5cm]{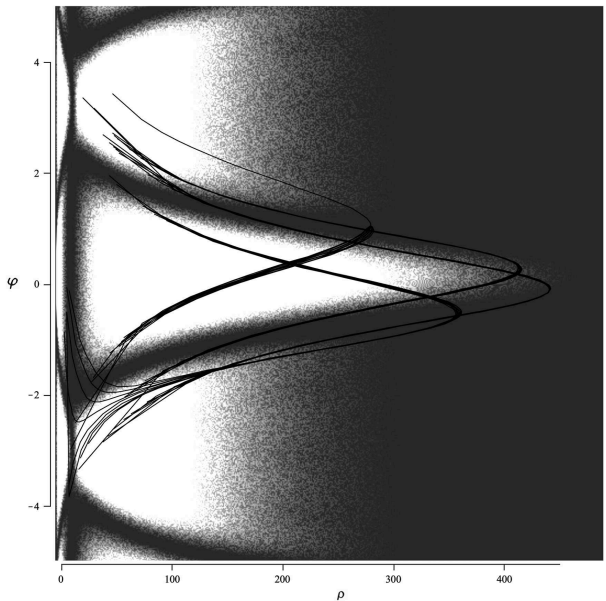}\\
  \caption{ {\small Density plot of $|\Psi|^{2}$ for a 6-term $\psi$ (n=-3,-2,-1,1,2,3) with equal amplitudes, $\theta=2.6\times10^{-5},~ \bar{\theta}=2.0\times10^{-5},~\lambda=0.193$ and $b=5\times10^{-4}$, which is shown to be remarkably consistent with the classical path.} } \label{Fig4}
\end{figure}

\section{Conclusions}

In this paper, we have revisited the issue of classical signature change in FRW cosmological model having a phase space with noncommutative collective coordinates, namely scale factor and scalar field, and their conjugate momenta. The conditions for which a classical continuous change of signature is possible, have been investigated. Comparison with the results of the commutative model studied in \cite{Dereli&Tucker} shows that the noncommutativity parameters affect the classical time evolution of both scale factor and the scalar field in this noncommutative signature changing cosmological model.

We have also studied the quantum cosmology of this noncommutative cosmological
model and obtained the corresponding solutions of Wheeler-DeWitt equation, perturbatively. Following the interesting issue of classical-quantum correspondence, we have shown that such correspondence is achieved in the noncommutative model of signature changing FRW cosmology, as well as the commutative model,
by use of ($\theta, \bar{\theta}, \lambda, b, m^2$) rather than ($\lambda, b, m^2$). In other words, noncommutativity not only does not destruct the classical-quantum correspondence but also helps for setting a better correspondence by using an enlarged parameter space.

The inequality (\ref{Eq51}) imposes a bound on $\lambda$ which shows the cosmological constant is negative. Demanding $\theta$ and $\bar{\theta}$ to be infinitesimal parameters, we find that (\ref{Eq51}) allows for a vast spectrum, from very small to very large values, for the negative bare cosmological constant. This is interesting because the bounds of cosmological constant are limited by the values of noncommutative parameters. This is important in that if noncommutativity really exists, then the approximate experimental values of the noncommutativity parameters may limit the experimental bounds on the cosmological constant.

Apart from the former two results mentioned in the conclusion, the latter result is of particular importance from high energy physics point of view. This is because both the cosmological constant and noncommutativity parameters play major roles in high energy physics and finding a relation like (\ref{Eq51})
motivates one to look for a general mechanism in which the cosmological constant problem is solved by the idea of noncommutativity.

It is also worth to mention that the solvability condition for Eq.(\ref{Eq27}), namely $l=0$, may be rewritten as

\begin{equation}\label{m}
m^{2}=\frac{1}{2}\,{\frac {3\,\bar{\theta}\,\lambda\,{\theta}^{2}+9\,{\lambda}^{2}{\theta}^ {3}+12\,\lambda\,\theta+4\,\bar{\theta}+16\,{b}^{2}{\theta}^{3}}{\theta\,
 \left( \bar{\theta}\,\theta+6\,\lambda\,{\theta}^{2}+4 \right) }},
\end{equation}
which asserts that: the physical parameters $(m^2,\lambda, b)$ are constrained
by the noncommutative parametres $(\theta,\bar{\theta})$. For example, one
may consider the mass of scalar field as an emergent parameter provided
that we consider $(\theta,\bar{\theta},\lambda, b)$ as given parameters. This is interesting because the mass may become tachyonic $m^2<0$, via some combinations of $(\lambda, b)$ and noncommutative parameters $(\theta,\bar{\theta})$. Actually, one may describe each of the parameters $(m^2,\lambda, b)$ in terms
of two remaining ones and noncommutative parameters, using (\ref{m}). So,
it is appealing to investigate if the above result is a general feature of the idea of noncommutativity in theories having a set of physical parameters.

\end{document}